\documentclass{aa}
\usepackage{psfig}
\begin{document}
    \title{Asiago eclipsing binaries program. III. V570 Per\thanks{based
          mainly on data obtained with Asiago 1.82 m telescope}$^{,}$\thanks{
          Table~1 available in electronic form only}}

\author	{L. Tomasella \inst{1}
	\and
	U. Munari\inst{1}
	\and
	S. Cassisi\inst{2}
	\and
	A. Siviero\inst{1}
	\and
	S. Dallaporta\inst{3}
        \and
        R. Sordo\inst{1}
	\and
	T. Zwitter\inst{4}
        }

\institute{
	INAF Osservatorio Astronomico di Padova, Sede di Asiago, 36012
	Asiago (VI), Italy
	\and
        INAF Osservatorio Astronomico di Collurania, Via M. Maggini, 64100 Teramo, Italy
	\and
	Via Filzi 9, I-38034 Cembra (TN), Italy
        \and
        University of Ljubljana, Department of Physics, Jadranska 19, 1000 Ljubljana, Slovenia
        }

   \date{Received date ................; accepted date ...............}

     \abstract{The orbit and physical parameters of the previously unsolved
     double-lined eclipsing binary V570~Per, discovered by the Hipparcos
     satellite, are derived using high resolution Echelle spectroscopy and
     $B$, $V$ photo-electric photometry. The metallicity from $\chi^2$
     analysis of the spectra is [M/H]=+0.02$\pm$0.03, and reddening from
     interstellar NaI and KI absorption lines is $E_{B-V}$=0.023$\pm$0.007. 
     V570~Per is a well detached system, with shallow eclipses due to low
     orbital inclination and no sign of chromospheric activity. The two
     components have masses of 1.449$\pm$0.006 and 1.350$\pm$0.006 M$_\odot$
     and spectral types F3 and F5, respectively. They are both still within
     the Main Sequence band (T$_1$=6842$\pm$25 K, T$_2$=6562$\pm$25 K from
     $\chi^2$ analysis, R$_1$=1.523$\pm$0.030, R$_2$=1.388$\pm$0.019
     R$_\odot$ derived forcing the orbital solution to conform to the
     spectroscopic light ratio) and are dynamically relaxed to co-rotation
     with the orbital motion ($V_{\rm rot} \sin i_{1,2}$=40 and 36 ($\pm$1)
     km~sec$^{-1}$).  The distance to V570~Per obtained from the orbital
     solution is 123$\pm$2~pc, in excellent agreement with the revised
     Hipparcos distance of 123$\pm$11~pc.  The observed properties of
     V570~Per components are compared to available families of stellar
     evolutionary tracks, and in particular to BaSTI models computed on
     purpose for exactly the observed masses and varied chemical
     compositions.  This system is interesting since both components have
     their masses in the range where the efficiency of convective core
     overshooting has to decrease with the total mass as a consequence of
     the decreasing size of the convective core during the central H-burning
     stage. Our numerical simulations show that, in order to match all empirical
     constraints including also the spectroscopical measurements, a small
     but not null overshooting is required, with efficiencies of $\lambda_{OV}$=0.14 and 0.11 for the
     1.449 and 1.350~M$_\odot$ components, respectively. This confirms the
     finding of Paper~II on the similar system V505~Per. At the
     $\approx$0.8~Gyr age of the system, the element diffusion has reduced the
     surface metallicity of the models from the initial [M/H]=+0.17 to
     [M/H]=+0.02, in perfect agreement with the spectroscopically derived
     [M/H]=+0.02$\pm$0.03 value.
     
     \keywords{stars: fundamental parameters --
                binaries: spectroscopic --
                binaries: eclipsing -- star: individual: V570 Per}
            }

   \maketitle

\section{Introduction}

In the present series of papers we derive global physical parameters for a
selection of double-lined eclipsing binaries (SB2 EBs) by means of Echelle
high resolution, high S/N spectroscopy and $B$, $V$ photoelectric
photometry. Orbital solutions provide masses and radii, reddening is
measured from intensity of interstellar absorption lines and atmospheric
analysis supplies effective temperatures, surface gravities, metallicity and
rotational velocities. The physical parameters we derive are of high quality
(1\% accuracy regime) and are used to constrain the input physics of
theoretical stellar models. We compute tailored evolutionary stellar models
for the exact masses and chemical mixtures observed in the components of the
binaries. This allows to better focus on the effect expected, for ex., from
overshooting and element diffusion.  Siviero et al. (2004) and Tomasella et
al. (2007), hereafter Paper I and Paper II, outline details of the type of
data and methods used throughout this program.

V570~Per (HD 19457, HIP 14673) is a nearby eclipsing binary of early F
spectral type, whose variability, characterized by a 1.9 days
period, was discovered by Hipparcos satellite. As it will be seen in the
next sections, its components do not show intrinsic variability. They are well
separated, perfectly round, slow rotating and well within their Roche lobes,
which make them proper tests for stellar models. V570~Per, though seen
projected toward the $\alpha$~Per (Melotte~20) young open cluster, 
is a foreground star not physically
associated to the cluster.  This can be deduced from the comparison between
the distance and proper motion of the cluster ($d$=183$\pm$7 pc,
$\mu_\alpha^*$=$+$22.47$\pm$0.16~mas yr$^{-1}$,
$\mu_\delta$=$-$25.99$\pm$0.17~mas yr$^{-1}$, from van Leeuwen 1999) and of
V570~Per ($d$=117$\pm$14 pc, $\mu_\alpha^*$=$+$52.20$\pm$0.85~mas yr$^{-1}$,
$\mu_\delta$=$-$41.58$\pm$0.79~mas yr$^{-1}$, from Hipparcos Catalogue).
The revised Hipparcos distance to V570~Per is $d$=123$\pm$11~pc (van Leeuwen
2007).

A preliminary photometric and spectroscopic study of V570~Per was presented
by Munari et al. (2001, hereafter M01). It was derived during an evaluation
of the performances expected from Gaia (an ESA Cornerstone mission) on
eclipsing binaries and, as such, the accuracy and abundance of input
observational data (Hipparcos/Tycho photometry and Gaia-like ground based
spectra) were necessarily lower than in the present study. A new study of
the binary based on a much improved set of observational data is therefore
justified, especially if supported by brand new reddening and atmospheric
analysis.

    \begin{figure*}
    \centerline{\psfig{file=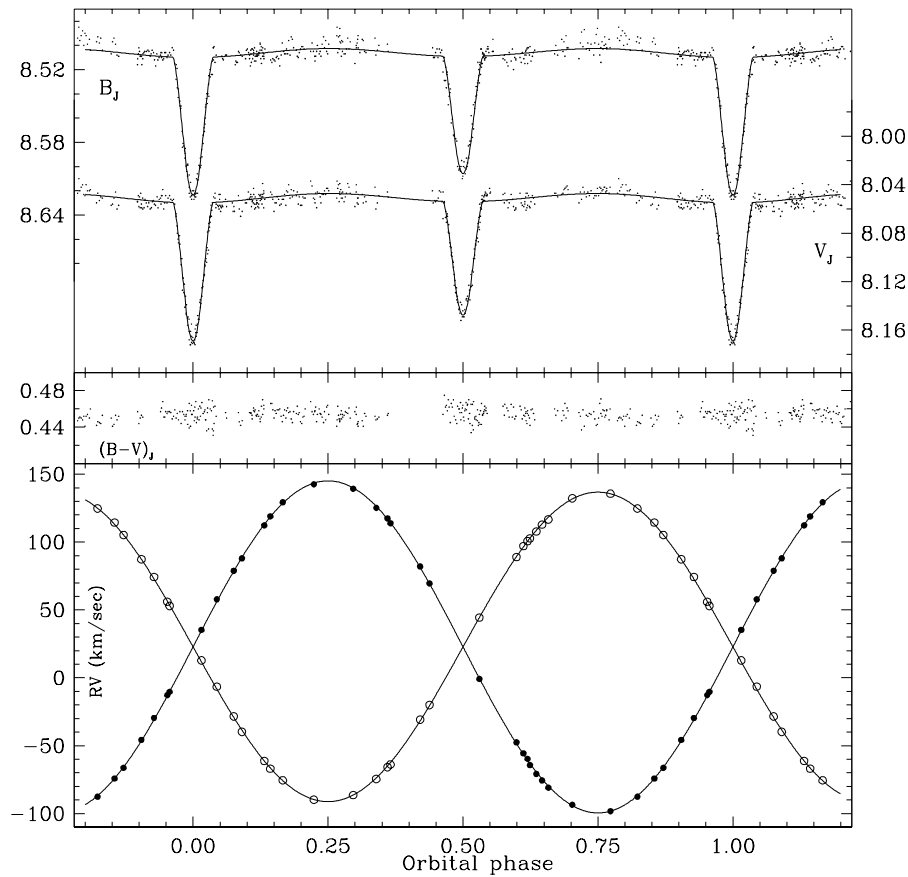,width=12.0cm}}
    \caption{The observed $B$, $V$, {\em B$-$V} and radial velocity curves of
	V570~Per. In the radial velocity panel, the open circles indicate the 
	hotter and more massive (primary) star, while the filled circles pertain to
	the cooler and less massive (secondary) star. The orbital solution
	from Table~3 is over-plotted to the observed data.}
    \end{figure*}

\section{Photometric data}

    \setcounter{table}{1}
    \begin{table}
    \caption{Measured radial velocities of V570~Per.}
    \begin{center}
    \centerline{\psfig{file=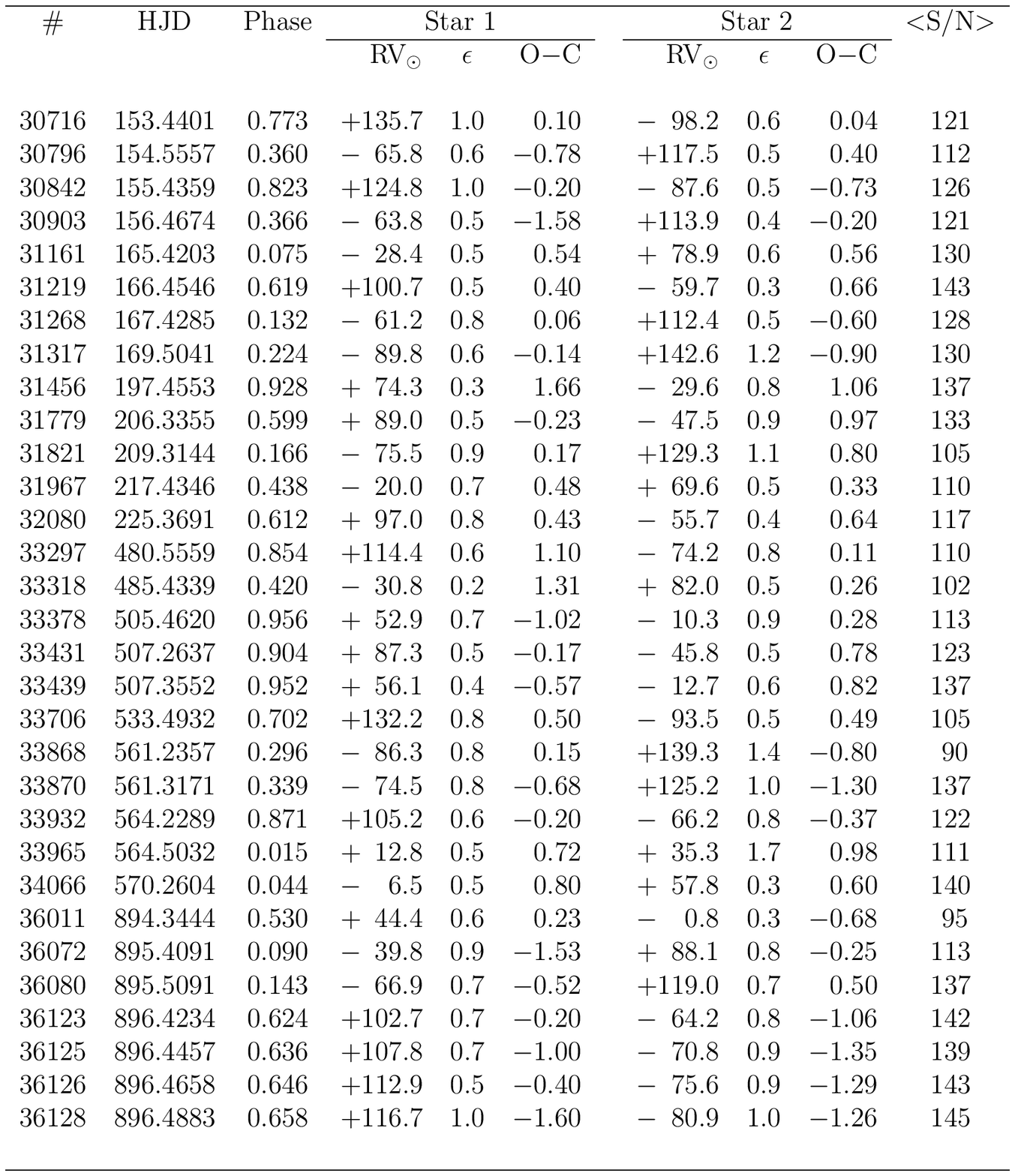,width=8.7cm}}
    \end{center}
    \end{table}

The photometric observations of V570~Per were obtained in $B$ and $V$
(standard Johnson filters) with a 28 cm Schmidt-Cassegrain telescope and an
Optec SSP5 photometer. The diaphragm has a diameter of 77 arcsec. There are no
stars in the aperture brighter than $V$=15.5 mag (contributing less than
0.001~mag to recorded photometry). The instrumentation already proved to be
very accurate and reliable (cf. Paper~I and Paper~II) and thus perfectly
suited to deal with the low amplitude eclipses of V570~Per (cf. Fig.~1).
 
The comparison star is HD 19805 (HIP 14980,  $B_{\rm T}$=8.108$\pm$0.015,
$V_{\rm T}$=7.973$\pm$0.012, spectral type B9.5~V) and the check star is
TYC~3315~308~1 ($B_{\rm T}$=9.919$\pm$0.029, $V_{\rm T}$=9.567$\pm$0.032). 
Both the comparison and check stars are close to V570~Per on the sky (the
distances being $\sim$42 and $\sim$18 arcmin respectively) so the
atmospheric corrections are rather small. All the observations were obtained
at heights over the horizon in excess of 30 degrees. The comparison star was
measured against the check star at least once every observing run. In all,
34 measurements of the magnitude difference comparison$-$check star were
collected. It remained constant through the whole observing campaign within
a standard deviation of 0.006 mag. This confirms Hipparcos findings that
both the comparison and the check stars are not variable, and thus well
suited to serve in the photometry of V570~Per. Following Bessell (2000)
transformations from Tycho to Johnson photometric system, we adopted for the
comparison star $B$=8.073 and $V$=7.957.

In all, 446 measurements in $B$ and 465 in $V$ were collected for V570~Per
between 2000 and 2003. They are listed in Table~1 (available
electronic only). Each photometric point is actually the mean of 10
consecutive and independent measurements (every one lasting 5 sec) and the
typical error of the mean is 0.006 mag in $B$ and
0.005 mag in $V$. All the observations are corrected for atmospheric
extinction and color equation via nightly calibration on Landolt's
equatorial fields.

The light curves of V570~Per in each band as well as the ({\em B$-$V}) color
are shown in Fig.~1. Spectroscopic and photometric observations are well distributed in phase.
The eclipses are very shallow due to the low inclination of the orbit
($i$=77$\degr$), and this is the main reason for the absence of
color variations during the eclipses (both stars remain essentially visible
throughout the whole eclipses). The mean brightness out of eclipses is
$B$=8.505 and $V$=8.053 mag.  The dispersion of the $B$ and $V$ measurements
about their mean value out of the eclipses ($\sigma_B$=0.007,
$\sigma_V$=0.006) is only marginally higher than the accuracy of a single
measurement. Thus any intrinsic variability of the amplitude larger than
0.007 mag should be ruled out.

\section{Spectroscopic observations}

The spectra of V570~Per were obtained in 1999$-$2002 with the Echelle+CCD
spectrograph on the 1.82 m telescope operated by INAF Osservatorio
Astronomico di Padova atop Mt. Ekar (Asiago). The instrumentation and
observing set-up exactly match those described in Paper I, to which we refer
for details of the observing mode. Here we recall that the wavelength region
covered is 4500$-$9480~\AA\, with a resolving power
$R$~$\sim$~20\,000. A journal of the observations is given in Table~2, in
which the first three columns list the spectrum number (from the Asiago Echelle log book), 
the heliocentric JD ($-$2451000) and the orbital phase, while the last 
column gives the S/N per pixel of the recorded stellar continuum  
averaged over the whole measured wavelength range (4890-5690~\AA). The 
other columns lists, for both components separately and in km~sec$^{-1}$, 
the radial velocity, its error and the difference with the computed orbit.  We 
obtained 31 spectra with exposure times ranging from 1200 to 1800~sec,
which guarantee a good S/N ratio (cf. Table~2) while avoiding smearing due
to the orbital motion (1500~sec correspond to less than 1\% in orbital
period).

As outlined in Paper~II, small residual spectrograph flexures are removed by
cross-correlating the rich telluric absorptions complexes at
5880$-$5940, 6275$-$6310, 6865$-$7050 and 7160$-$7330~\AA\ with a
synthetic telluric absorption spectrum. Night-sky lines ([OI] and OH) and
city lines (HgI and NaI) are used to check the accuracy of the
wavelength calibration. Their mean velocity was found to be 0.0$\pm$0.1
km~sec$^{-1}$ on every spectrum.

\subsection{Radial velocities}

In deriving the radial velocities of V570~Per we followed strictly the
method outlined in details in Paper I. In short, six adjacent Echelle
orders, covering with no inter-order gaps the 4890$-$5690~\AA\ 
wavelength range, are measured via two-dimensional
cross-correlation technique (TODCOR), based on Zucker \& Mazeh (1994) algorithm. These
six Echelle orders are chosen because they are densely populated by
absorption lines, in particular FeI and MgI, that perform particularly well
in terms of radial velocity. As templates we use synthetic spectra with the
appropriate temperatures, surface gravities and rotational velocities,
selected from the Munari et al. (2005, hereafter M05) synthetic spectral atlas computed at
the same 20\,000 resolving power as the Echelle scientific spectra. The high
reliability of our TODCOR-based radial velocities has been subjected to
extensive tests described in Paper~II. Repeated here on V570~Per data, they
performed equally well, and we can conclude that no {\em external}
systematic error in excess of 0.25~km~sec$^{-1}$ affect our radial
velocities.

The results of the radial velocity measurements are reported in Table~2.
The mean error of radial velocities is 0.6~km~sec$^{-1}$ for star~1, and
0.7~km~sec$^{-1}$ for star~2, as estimated from comparison of the radial
velocities obtained separately from each of the six Echelle orders analysed
here.

\subsection{Reddening}

The measurement of reddening is a key step in the determination of the
absolute temperature scale (and therefore of the distance) of eclipsing
binaries. In spite of its short distance, some reddening is expected to
affect V570~Per given its low galactic latitude ($l$=145.18, $b$=$-$8.19).

Our spectra cover the interstellar NaI (5890 and 5896 \AA) and KI (7665 and 
7699 \AA) doublets which are excellent estimators of the reddening as
demonstrated by Munari \& Zwitter (1997). They calibrated a tight relation
linking the NaI D2 (5890~\AA) and KI (7699~\AA) equivalent widths with the
$E_{B-V}$ reddening. On spectra obtained at quadratures, lines from both
components are un-blended with the interstellar ones, which can be therefore
accurately measured. An example of such a spectrum is presented in Fig.~2.
We derived and equivalent width of 0.08$\pm$0.03~\AA\ for NaI~D2~(5890 \AA),
which corresponds to $E_{B-V}$=0.023$\pm$0.007 mag. This is the value
adopted in the rest of the paper. At such a low reddening, no detectable KI
interstellar line is expected, as confirmed by our spectra.

   \begin{figure}
    \centerline{\psfig{file=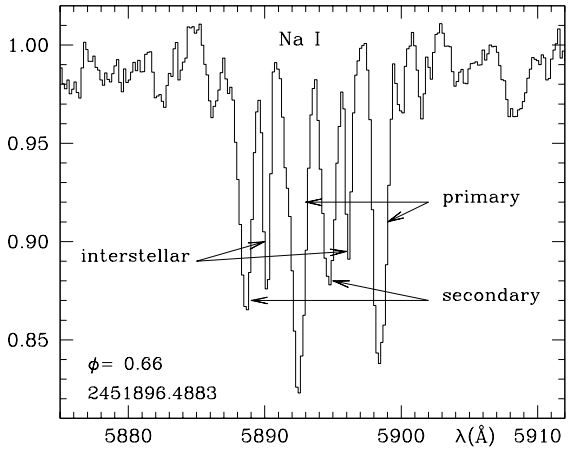,width=8.0cm}}
    \caption{The NaI doublet (5890, 5896 \AA) region for V570~Per. The 
	interstellar components are well separated from the stellar ones.}
    \end{figure}

As mentioned above, V570~Per is seen in the foreground of $\alpha$~Per
cluster. The cluster extends over a wide area and suffers from a patchy
distribution of the reddening, which amount to $E_{B-V}$=0.09 according to
Crawford \& Barnes (1974), and $E_{B-V}$=0.11 following Prosser (1992). It
is worth noticing that, according to Perry \& Johnston (1982), HD~19665, which
lies less than 1 arcmin from V570~Per, is affected by $E_{B-V}$=0.022
reddening.

\section{Orbital solution}

   \begin{table}
    \caption{Orbital solution for V570~Per. The set of parameters in column~2
    refer to the unconstrained solution described in Sect.~4, those in
    column~3 to the luminosity-constrained solution of Sect.~6.  Formal
    errors to the solutions are given. $d_{\rm orb}$, $d_{\rm HIP1997}$ and
    $d_{\rm HIP2007}$ are the distances derived by our orbital analysis,
    the original Hipparcos parallax (ESA 1997) and its recently revised
    value (van Leeuwen 2007), respectively.}
    \begin{center}
    \centerline{\psfig{file=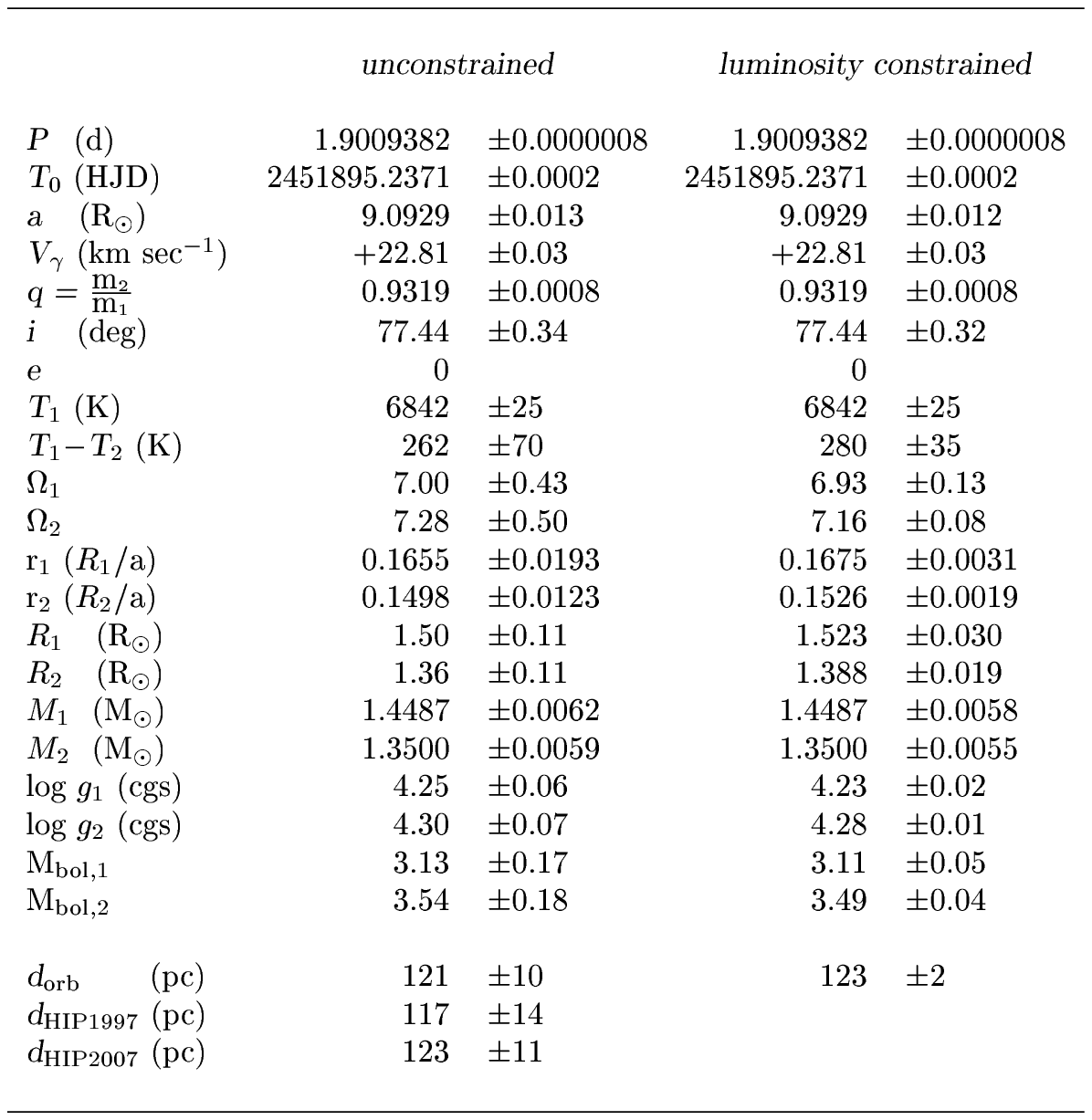,width=8.9cm}}
    \end{center}
    \end{table}

\begin{table}
    \caption{Atmospheric parameters (and their error of the mean, cf. Sect.~5)
    of V570~Per from the $\chi^2$ fit to synthetic spectra. They are
    compared to the results from orbital solution for the elements in common
    ($\Delta T_{\rm eff}$, $\log g$, $V_{\rm rot} \sin i$).}
    \begin{center}
    \centerline{\psfig{file=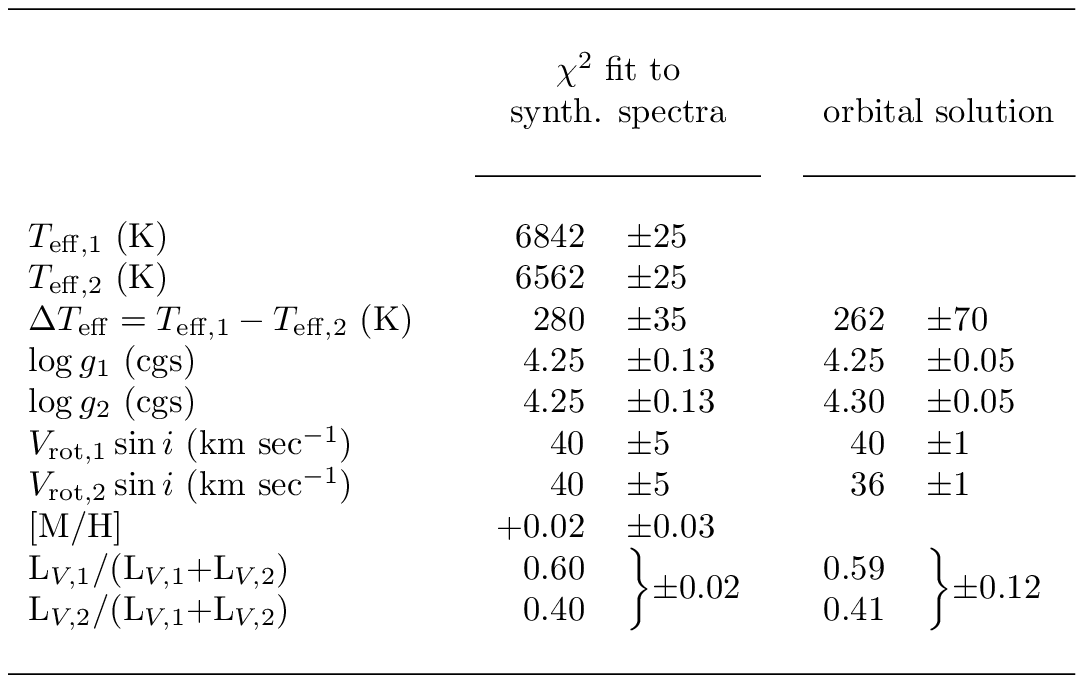,width=8.5cm}}
    \end{center}
    \end{table}

The orbital modeling is performed with version $WD98K93$ (Milone et al.
1992) of the Wilson-Devinney code (Wilson \& Devinney 1971, Wilson 1998),
which incorporates modified stellar atmospheres. The linear limb darkening
coefficients are taken from van Hamme (1993) for the appropriate
$T_{\rm eff}$, $\log g$, [M/H]. The temperature of the
primary adopted in the orbital solution, $T_{\rm 1}$=6842$\pm$25 K, is that
determined by atmospheric analysis (see Sect.~5 below). It is in close 
agreement with indications from spectral classification and de-reddened colors. 
Comparing the spectrum of V570~Per in the 8480$-$8740~\AA\ 
wavelength region with the Gaia spectral atlas of Munari \& Tomasella
(1999, which was compiled from observations of MK standards obtained with
the same instrument and resolution of the V570~Per spectra discussed in this 
paper), a spectral classification F3V+F5V is derived. The temperature
for the F3V primary, according to Bertone et al. (2004) calibration, is
6845~K. The color out of eclipse is {\em B$-$V}=+0.452 ($\pm$0.005), 
that corrected for $E_{B-V}$=0.023 ($\pm$0.007), provides ({\em
B$-$V})$_{\circ}$=+0.429 ($\pm$0.009). Comparing with Drilling \& Landolt (2000)
color calibration for main sequence stars, it corresponds to a spectral type
F4V, i.e. the mean spectral type of the two components of the binary. 

    \begin{figure*}
    \centerline{\psfig{file=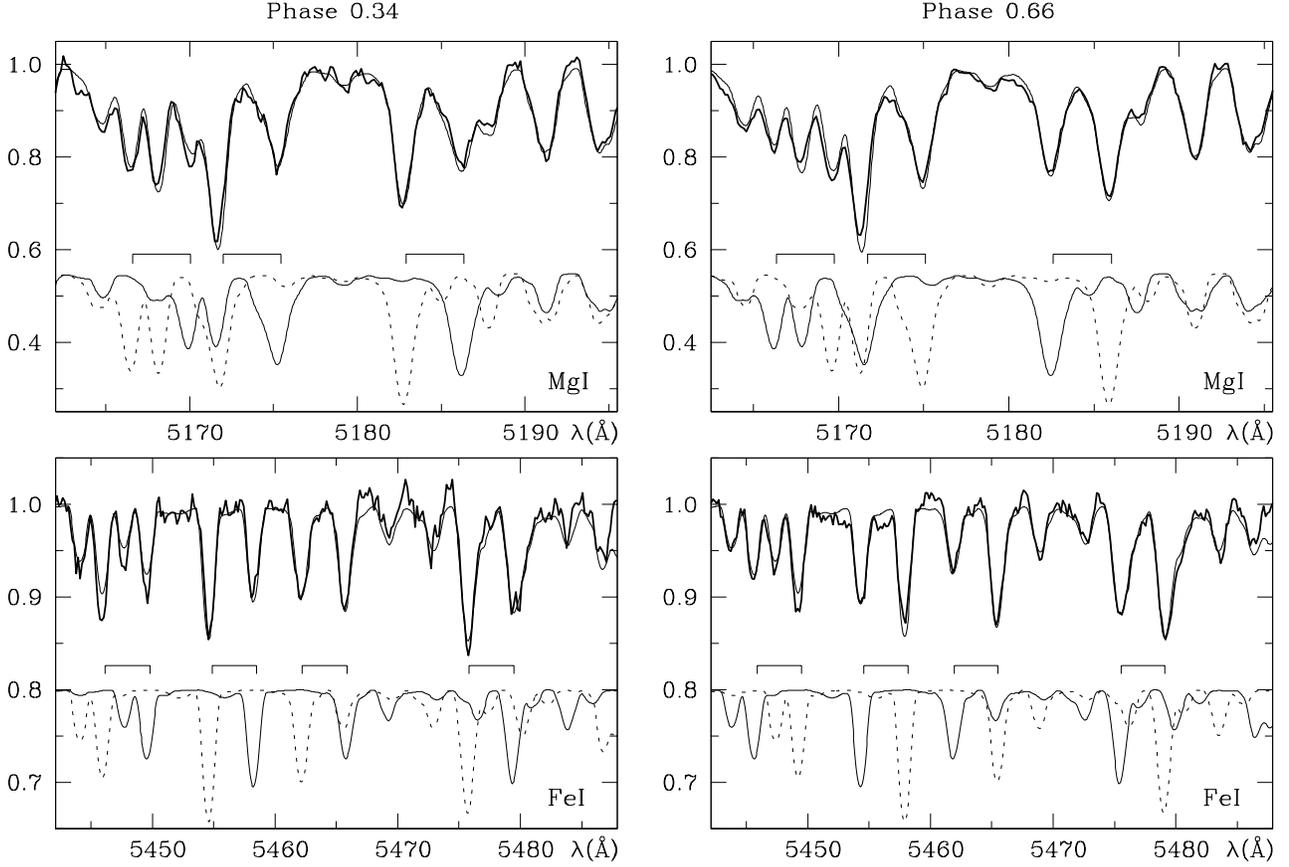,width=17.6cm}}
    \caption{Comparison between observed (thick line) and synthetic (thin line)
    V570~Per spectra over two sample wavelength regions dominated by MgI lines
    (top) and FeI lines (bottom). Spectra at orbital phases 0.34 and 0.66 (\#
    33870 and 36128 in Table~2, respectively) are shown, which allows full
    and reversed splitting of the absorption lines of the two components.
    In each panel the lower curves represent (not to scale with the
    main spectra but in correct proportion between them) the contribution of
    each component of the binary to the formation of the observed spectrum at
    the given phase. The markers connect the shifted wavelengths of the same
    MgI (top panels) and FeI (bottom panels) lines in the spectra of the
    two components of the binary.}
    \end{figure*}

The Gaia-like orbital solution of M01 was adopted as the set of initial
values for input parameters in the orbital modeling (period $P$, epoch
$T_{\rm 0}$, semi-major axis $a$, barycentric radial velocity $V_{\rm
\gamma}$, mass ratio $q=M_1/M_2$, inclination $i$, eccentricity $e$,
modified Kopal potentials $\Omega_{\rm 1,2}$ and $L_1/L_2$ ratio in each
pass-band).  

We computed the solution using mode 2 of the Wilson-Devinney code, which is
appropriate for detached binaries with no constraints on the potentials. 
The adopted bolometric albedos and exponents in the bolometric gravity
brightening law were set to $A_{1}$=$A_{2}$=0.5 and $g_{1}$=$g_{2}$=0.3,
respectively, as appropriate for convective atmospheres (Wilson 1998). As in
Paper~II, full orbital solution runs were carried out also with logarithmic
and square-root limb darkening laws, as well as with various combinations of
$A_{1,2}$ and $g_{1,2}$ in the range $0.5\leq$$A_{1,2}$$\leq 1.0$ and
$0.3\leq$$g_{1,2}$$\leq 1.0$ ($A$ and $g$ are expected to be of the order of
unity for radiative envelopes). No improvement in the formal accuracy of the
solution was obtained. No orbital parameter was found to vary, as a results
of these extensive experiments, by more than its formal error listed in the
final orbital solution of Table~3. The latter was obtain with the following
parameters for the linear limb darkening law: $x_{bol,1}$=0.607,
$x_{bol,2}$=0.664, $x_{V,1}$=0.407, $x_{V,2}$=0.428, $x_{B,1}$=0.504,
$x_{B,2}$=0.525. No evidence was found for multiple reflection effects
or third light presence.

The orbital period of V570~Per is stable. Our photometry provides five
epochs of minima (all of them with an uncertainty of $\pm$0.00012 days):
primary eclipses on HJD=2451895.23713, 2451920.44464, 2452986.37566 and
secondary ones on HJD=2451951.31480, 2452158.51707. The Hipparcos and Tycho
Catalogues (ESA 1997) give 2448500.1520$\pm$0.001 as the time of primary
minimum, revised to 2448500.1655$\pm$0.003 in M01 re-analysis in combination
with radial velocities. All these minima are well fitted by the
1.9009382$\pm$0.0000008 day period of the orbital solution in Table~3.

V570~Per light and radial velocity curves do not show signature of
an eccentric orbit. The circularity of the orbit is confirmed by initial
modeling runs during which eccentricity was allowed to vary and nevertheless
it did not drift away from zero. After a few such trials $e$ was set to zero
for the rest of orbital modeling.

The orbital solution of Table~3 is over-plotted to the observational data in
Fig.~1. Their r.m.s is 0.006 mag for both $B$ and $V$ photometry, 0.8 km sec$^{-1}$
for radial velocities. 
The formal accuracy is 0.41\% and 0.44\% on the mass of the primary and
secondary, respectively, and 7.3\% and 8.1\% on their radii. The latter 
is inflated by the low orbital inclination of the system ($i$=77.44$^{\circ}$
$\pm$0.34), that produces grazing eclipses of low amplitude 
($\Delta B_{I}$=0.116, $\Delta B_{II}$=0.094, $\Delta V_{I}$=0.117,
$\Delta V_{II}$=0.098 mag).

As a test, separate orbital solutions were carried out combining in turn
$B$ or $V$ photometry alone with the radial velocities.
We converged to the same orbital parameters of Table~3, with larger formal
errors, as expected. Furthermore, the orbital solution was carried out on $B$ and
$V$ photometric data together, without including radial velocities in the
fitting process, and retaining $i$, $M_{1}$, $M_{2}$ from the final solution
of Table~3. Again, the orbital solution converged - within the formal errors
- to the same set of parameters given in Table~3, with
$R_1$=1.55$\pm$0.02~$R_\odot$ (instead of
1.50$\pm$0.11) and $R_2$=1.33$\pm$0.02~$R_\odot$ (instead of 1.36$\pm$0.11).

To check the accuracy of the derived orbital solution, we compared the
corresponding distance with the Hipparcos parallax.  In calculating the
distance, we adopted from Bessell, Castelli and Plez (1998) a bolometric
magnitude $M_{\rm bol,\odot}$=4.74 for the Sun and, for the corresponding
atmospheric parameters, a bolometric correction $BC$=0.00 for both V570~Per
components. The resulting distance, for the $E_{B-V}$=0.023 color excess
derived in sect~3.2, is $d$=121$\pm$10 pc, in agreement with the
123$\pm$11 pc revised Hipparcos parallax for V570~Per (van Leeuwen 2007 and
private communication in advance of publication).

Varying the adopted value for the solar bolometric magnitude (e.g. from
$M_{\rm bol,\odot}$=4.72 reported by Straizys and Kuriliene 1981, to $M_{\rm
bol,\odot}$=4.74 by Livingston 2000, and $M_{\rm bol,\odot}$=4.83 by Popper
1980) and the corresponding bolometric correction for F5V stars
($BC$=$-$0.02, $-$0.08 and $-$0.03, respectively), the distance derived from
the orbital solution changes by $\pm$ 2 pc.

\section{Atmospheric analysis}

To derive the atmospheric parameters of both V570~Per components, we
performed a $\chi^2$ fit against the synthetic spectral atlas of M05.
It covers the 2500--10\,500 \AA\ wavelength range at the same resolving power
of our spectra ($R~$=~20\,000). The M05 synthetic spectra are calculated with the 
revised solar abundances by Grevesse \& Sauval (1998) and the new opacity
distribution functions (ODFs) of Castelli \& Kurucz (2004),
throughout the whole HR diagram for 12 different rotational velocities, $T_{\rm eff}$
ranging from 3500 to 47\,500 K, $\log g$ from 0.0 to 5.0, [M/H] from $-$2.5
to +0.5, two values of
$\alpha$-enhancement ([$\alpha$/Fe]=0.0 and +0.4), and three 
micro-turbulent velocities (1, 2, 4 km~sec$^{-1}$).
The $\chi^2$ fitting was restricted to the higher S/N spectra obtained at
both quadratures. We limited the fitting to the same 4890--5590 \AA\ wavelength range
used to derive the radial velocities. 

In the $\chi^2$ fitting procedure we fixed only the radial velocities of the
two components, while $T_{\rm eff}$, $\log g$, $V_{\rm rot}$~$\sin i$,
[M/H], micro-turbulent velocity and brightness ratio for both components
were treated as free parameters. The results are given in Table~4. The
listed errors are the error of the mean computed on the results obtained
separately on the six Echelle orders covering the 4890--5590 \AA\ wavelength
range. Table~4 also provides a comparison with parameters in common with
the orbital solution, and it fosters strong confidence in the results of the
atmospheric analysis. Figure~3 highlights the goodness of the match between
observed and synthetic spectra by over-plotting sample wavelength intervals
of them.

\begin{figure} 
\centerline{\psfig{file=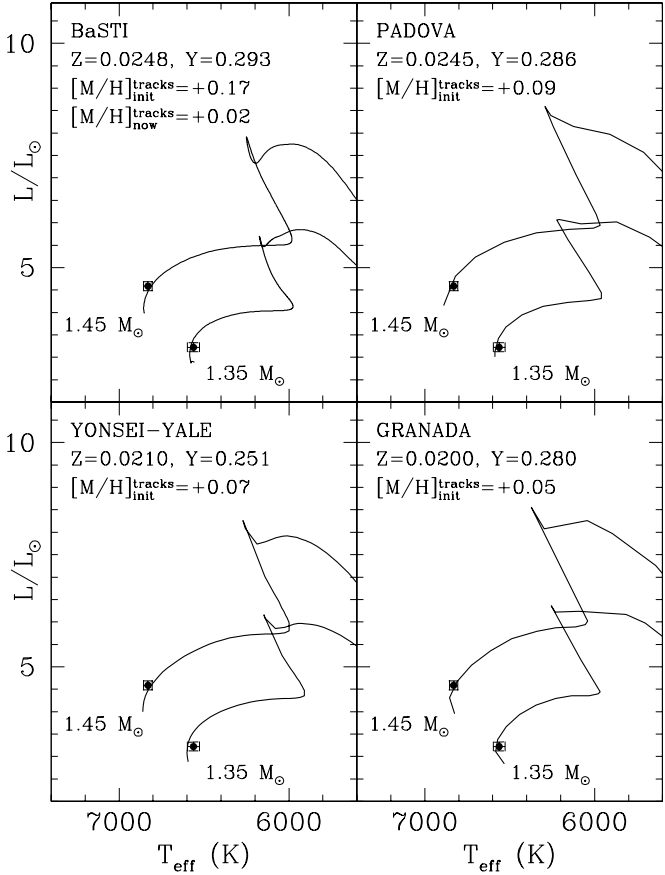,width=8.8cm,angle=270}}
\caption{Comparison of evolutionary tracks with observed parameters for
 V570~Per components (filled circles with error bars).  The metallicity,
 helium content and mass of the fitting tracks are given in each panel.
 Padova, Yonsei-Yale and Granada tracks are interpolated from published
 ones, while BaSTI tracks have been computed on purpose for this work for
 the exact masses of the two components (1.4487 and 1.3500 M$_{\odot}$).}
\end{figure}

In a configuration such as that of V570~Per, with similar stars and
partial eclipses that are relatively shallow, the information on the
relative sizes of the stars is weak. This reflects in the appreciable
uncertainties of the radii in Table~3. The brightness ratio of the two stars
as determined by the $\chi^2$ fit offers an important and independent check
on the accuracy of the radii. Such comparison is provided in Table~4 as
brightness ratio of the two stars. It is carried out in the $V$ band because
at its effective wavelength are centered the six contiguous echelle orders
used in our atmospheric analysis. The brightness ratios derived by $\chi^2$
fit and orbital solution are the same, well within the respective errors.
Also the TODCOR algorithm converged on the same 0.60/0.40 brightness ratio
when deriving the radial velocities of the two components.

The V570~Per components rotate in synchronicity with the orbital motion (as
highlighted by the comparison between observed and expected $V_{\rm rot 1,2}
\sin i$ in Table~4), have spherical shapes ($R_{\rm L1}=R_{\rm
pole}$) and are widely separated ($R$/$a$~$\sim$~0.16). Under these
circumstances, no tidal-induced chromospheric activity is expected. This is
confirmed by the absence of emission line cores or emission veiling in the
CaII~H \& K blue doublet or CaII far-red triplet lines, as verified at all
orbital phases by comparison with synthetic spectra.

\section{Constraining the stellar radii}

The shallow eclipses displayed by V570~Per cannot sharply constrain the
stellar radii, as indicated by the $\sim$8\% error on the radii obtained in
the unforced orbital solution of Sect.~4, which is listed in the second
column of Table~3. However, a relevant constrain is provided by the
luminosity ratio of the two components as determined by $\chi^2$ analysis
and TODCOR fitting (cf Sect.~5 and last two lines of Table~4). We have
therefore re-run the orbital solution with the Wilson-Devinney code,
operated this time in mode~0, which is specific to the case when the
luminosities are externally supplied. We then rapidly converged to the
luminosity-constrained obital solution listed in the third column of
Table~3. It confirms the unforced orbital solution of Sect.~4, with an
overall decrease of formal errors and in particular of those on stellar
radii and on the distance to the binary system. The radii improved from
1.50$\pm$0.11, 1.36$\pm$0.11 R$_\odot$ to 1.523$\pm$0.030, 1.388$\pm$0.019
R$_\odot$, and the distance to the system from 121$\pm$10 to 123$\pm$2~pc.
The latter is now identical to the revised Hipparcos distance (123$\pm$11~pc,
van Leeuwen 2007).

The comparison of V570~Per with theoretical stellar models carried out in
next Sect.~7 and in Figures 4-6 is based on this luminosity-constrained obital
solution.

\section{Observations vs theory}

As it has been extensively reviewed in Paper II, binary systems with
well-measured masses, radii, effective temperatures and heavy elements
abundance provide stringent tests for evolutionary stellar models.  Many
issues can be investigated by comparing empirical evidence for the binary
stars and theoretical predictions such as the efficiency of super-adiabatic
convection in low-mass stars, the efficiency of diffusive processes (atomic
diffusion and radiative levitation) and the real extension of convective
core overshooting during the central H-burning stage. For intermediate mass
stars, i.e. stars more massive than about 1.2 M$_\odot$, this latter issue
is really a long-standing problem (see e.g. Cassisi 2004 and references
therein).  An additional, still unsettled issue is how convective core
overshooting reduces in response to a decrease in stellar mass and its
canonical convective core (see Woo \& Demarque 2001 for a detailed
discussion).

\begin{figure}
\centerline{\psfig{file=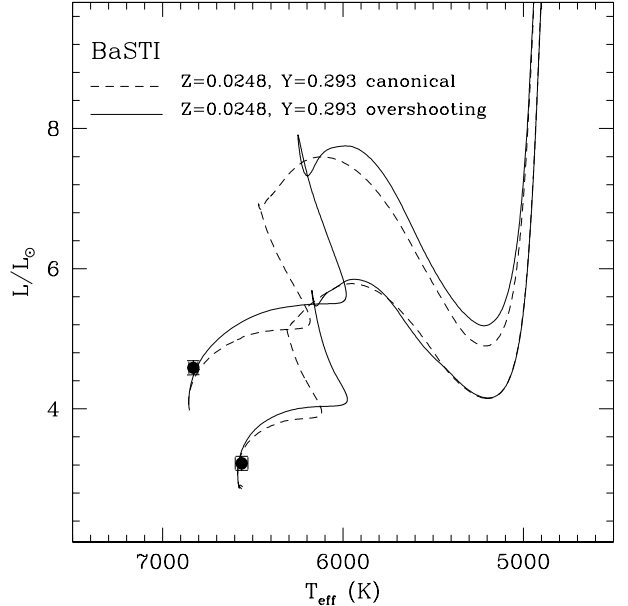,width=8.1cm}}
\caption{Comparison of V570~Per components with BaSTI tracks computed - for
         the exact observed masses - with and without overshooting (see
         Sect.~6 for details).}
\end{figure}

\begin{figure}
\centerline{\psfig{file=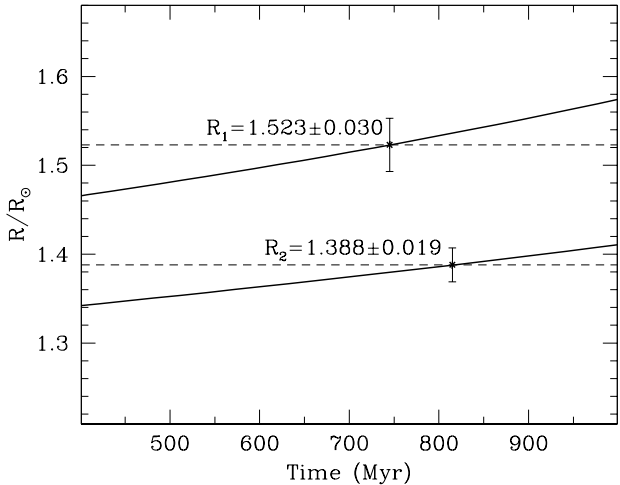,width=8.2cm,angle=0}}
\caption{Comparison between the observed radii for the V570~Per components
(dashed lines), their uncertainty (error bars) and the BaSTI evolutionary 
tracks (solid lines) computed with overshooting and for exactly the observed 
individual masses. The resulting age for V570~Per is $\approx$790$\pm$60~Myr.}
\end{figure}

In Paper II, we have briefly outlined the main results that have been
obtained in this context by using empirical data for galactic Open Clusters
of suitable age. We noticed the strong limitations imposed by the
uncertainties in the distance, reddening and heavy elements abundances
adopted for the clusters (cf. Vandenberg et al. 2007).  A firmer
contribution can be provided by eclipsing binary systems of suitable masses,
with examples provided by V459~Cas (Sandberg Lacy et al. 2004), TZ~For
(Vandenberg et al. 2006), AI~Phe (Andersen et al. 1988, Pietrinferni et al.
2004, hereafter P04) and V505~Per (Paper II). The above derived
1.449$\pm$0.006~M$_\odot$ and 1.350$\pm$0.006~M$_\odot$ masses for the two
components of V570~Per qualify this binary too as a suitable constrain to
stellar theoretical models and worth a detailed discussion.

V570~Per measured properties have been compared with the following sets of
theoretical stellar models: BaSTI (P04, see also Cordier et al. 2007), 
Padova (Girardi et al. 2000), Yonsei-Yale (Yi et al. 2001, second release), and Granada 
(Claret et al. 2003), which have all been computed by assuming a
scaled-solar heavy elements distribution. 

The loci of theoretical isochrones and of V570~Per components on the HR
diagram are shown in Fig.~4. For Yonsei-Yale, Granada and Padova models, we
interpolated linearly both in mass and metallicity between the closest
published isochrones. BaSTI models, on the other hand, have been computed on
purpose for this paper, for exactly the measured masses of the two V570~Per
components and exploring a fine grid in metal abundance, helium content,
element diffusion, and amount of core overshooting (including canonical
models with no overshooting). The overshooting was treated in the same way
as in Paper~II, and as extensively discussed by P04.

Padova and Granada models do not account for elements diffusion during the
life of a star, and Yonsei-Yale does it only for helium. Thus, the
metallicity of their fitting isochrones in Fig.~4 pertains to the initial
composition when V570~Per was born, and not to its present value at the
stellar surface. It is therefore expected that the metallicities derived
by fitting to Padova, Granada and Yonsei-Yale isochrones are {\em larger}
than measured by spectroscopy. The best fitting BaSTI models have an 
initial metallicity [M/H]=+0.17 for both stars, that decreases to
[M/H]=+0.00 and +0.04 for the 1.45 and 1.35 M$_\odot$ components, respectively, 
by the time of V570~Per current 0.8~Gyr age (see below for its derivation).
The average of this values very well compares with the spectroscopically
measure of [M/H]=+0.02$\pm$0.03 for the V570~Per components.

The comparison with theoretical isochrones in Fig.~4 shows that both
components of V570~Per are only marginally evolved. The more
massive of the two stars has already moved slightly away from the zero age
main sequence. Figure~5 shows the effect of overshooting, which has
been computed for BaSTI models according to the same P04 prescriptions
already adopted in Paper~II. This figure shows that canonical stellar models 
with no overshooting but accounting for element diffusion, can still reproduce 
the position of V570~Per in the H-R diagram (dashed line in Fig.~5). 
However, the canonical models give a surface metallicity at the present age of the binary system
equal to [M/H]=+0.12, which is outside the range of uncertainty for the observed
[M/H]=+0.02$\pm$0.03 value. 
On the other hand, a
small amount of overshooting allows a perfect fit for both the spectroscopic
metallicity and the H-R diagram location, as illustrated by the solid lines in Fig.~5 that have been
computed for an efficiency of the overshooting amounting to $\lambda_{\rm
OV}$=0.14 for the 1.449$\pm$0.006~M$_\odot$ primary and $\lambda_{\rm
OV}$=0.11 for the 1.350$\pm$0.006~M$_\odot$ secondary. In Paper II, for the
lighter components of V505~Per, we obtained proportionally lower
efficiencies: $\lambda_{\rm OV}$=0.093 and 0.087 for  
the 1.269 and 1.251 M$_{\odot}$ components, respectively.
The efficiency is
usually defined in terms of the parameter $\lambda_{\rm OV}$ that gives the
length - expressed as a fraction of the local pressure scale height $H_{\rm
P}$ - crossed by the convective cells in the convective stable region
outside the Schwarzschild convective boundary. Regardless
of the initial metallicity, P04 adopt $\lambda_{\rm OV}$=0.20$\times$$H_{\rm
P}$ for $M$$\geq$1.7~M$_\odot$, $\lambda_{\rm OV}$=0 for 
$M$$\leq$1.1~M$_\odot$, and $\lambda_{\rm OV}$=($\frac{M}{M_\odot}$ $-$
0.9)/4 for 1.1 M$_\odot\leq$M$\leq$ 1.7 M$_\odot$. 

To estimate the age of the system, in Fig.~6 we plot the evolution in radius
of the BaSTI models above described for the two V570~Per components and
compare them with the observed values. The derived system age is 790$\pm$60~Myr.

\begin{acknowledgements}
We would like to thank the anonymous referee for valuable suggestions, R.
Barbon and E. Nasi for useful discussion, F. van Leeuwen who kindly
communicated the revised Hipparcos parallax of V570~Per in advance of
publication, and the technical staff operating the 1.82m telescope in Asiago
for the skillful assistance during the whole project.
\end{acknowledgements}

\end{document}